# Parity-Time Symmetry and Exceptional points: A Tutorial

Alex Krasnok[1], Nikita Nefedkin[1], Andrea Alú[1,2]

[1]*Photonics Initiative, Advanced Science Research Center, City University of New York, New York, NY 10031, USA*

[2]*Physics Program, Graduate Center, City University of New York, New York, NY 10016, USA*

## Abstract

The physics of systems that cannot be described by a Hermitian Hamiltonian, has been attracting a great deal of attention in recent years, motivated by their nontrivial responses and by a plethora of applications for sensing, lasing, energy transfer/harvesting, topology and quantum networks. Electromagnetics is an inherently non-Hermitian research area because all materials are lossy, loss and gain distributions can be controlled with various mechanisms, and the underlying systems are open to radiation. Therefore, the recent developments in non-Hermitian physics offer exciting opportunities for a broad range of basic research and engineering applications relevant to the antennas and propagation community. In this work, we offer a tutorial geared at introducing the unusual electromagnetic phenomena emerging in non-Hermitian systems, with particular emphasis on a sub-class of these systems that obey parity-time (PT) symmetry. We discuss the basic concepts behind this topic and explore their implications for various phenomena. We first discuss the basic features of $\hat{P}$, $\hat{T}$ and $\hat{P}\hat{T}$ operators applied to electromagnetic and quantum mechanical phenomena. We then discuss the exotic response of PT-symmetric electromagnetic structures and their opportunities, with particular attention to singularities, known as exceptional points, emerging in these systems, and their unusual scattering response.

## 1 Introduction

Since the inception of the Dirac-von Neumann formulation of quantum mechanics, it has been assumed that quantum mechanical systems should be described by self-adjoint or Hermitian operators, i.e., operators invariant upon Hermitian conjugation, $\hat{O} = \hat{O}^{\dagger}$ (transposition plus complex conjugation) on a Hilbert space, because these operators correspond to observable real-valued physical quantities [1]. Indeed, observable quantities are always real-valued, whereas non-Hermitian systems are known to lead to complex eigenvalues and lack of time invariance of the wave function norm. However, in electromagnetics we are used to systems that do not obey Hermiticity and power conservation. Loss and gain in materials, as well as radiation loss in an open resonator, are classic examples of phenomena that lead to non-Hermitian responses.

The seminal work of Bender and his colleagues [2], [3] has somehow reconciled, or at least connected, these two camps, showing that Hermeticity is not a necessary condition to yield a real spectrum of eigenvalues. More specifically, Bender and Boettcher have found that non-Hermitian Hamiltonians can be found to possess a real-valued energy spectrum, and therefore in principle, may support observable phenomena in quantum mechanics. These Hamiltonians obey a special form of symmetry, known as parity-time (PT) symmetry [2], [4]–[7]. The pseudo-Hermiticity induced by PT-symmetry has drawn curiosity in various fields, producing exotic predictions in quantum field theory [3], Lie algebras [8] and complex crystals [9]. However, experimental demonstrations of the new physics of non-Hermitian Hamiltonians are deemed extremely challenging, if not impossible, in the quantum realm. On the other hand, electromagnetics offers a powerful platform for a more straightforward implementation of non-Hermitian phenomena. Because of the mathematical equivalence between the single-particle Schrodinger equation and the paraxial electromagnetic wave equation, the first proposed implementation of these concepts in electromagnetics was based on optical waveguides, which, through precise control of the gain and loss distribution in space, enabled the observation of PT-symmetric phenomena in classical settings [10]–[13]. Since then, there has been an ever-growing interest in exploring PT-symmetry, extended to many other domains, including atomic systems [14]–[16], quantum physics [17], electronics [18], and acoustics [19]–[21]. It has also been recognized that certain features of PT-symmetric systems may be realized in passive systems [22]–[24], with temporal modulations [25] or with non-monochromatic excitations [26], enabling easier implementations of these concepts that do not require a careful control of the distribution of optical gain, and facilitating the broader exploration of exotic non-Hermitian phenomena.

One of the most exciting effects associated with PT-symmetry is their phase transition, arising from spontaneous breaking around the so-called exceptional points (EPs). At these transitions, the eigenvalue spectrum of these systems from real-valued becomes complex [2],[3],[27]–[31], unveiling its non-Hermitian nature. EPs are special points in the parameter space of a non-Hermitian system at which the eigenvalues and corresponding eigenvectors simultaneously coalesce, resulting in their degeneracy. The abrupt nature of this phase transition leads to intriguing phenomena, including enhanced sensing [32]–[42], exotic lasing phenomena [43]–[46], optical isolation and nonreciprocity [47]–[50], loss-induced transparency [24], unidirectional invisibility [51]–[54], band merging [55], [56], robust wireless power transfer [57],[58], and topological chirality [59], [60].

In this work, we overview PT-symmetry and exceptional points, with a particular emphasis on their opportunities for electromagnetic systems, antennas and propagation. We start by considering the general concepts underlying this research area in quantum physics and photonics. Then, we classify PT-symmetric systems and discuss their basic features, emphasizing anomalies in electromagnetic scattering emerging in these systems and the role of PT-symmetry and exceptional points in these phenomena.

## 2 P, T, and PT Symmetry

Symmetries play a crucial role in field theories, such as in electromagnetism and quantum mechanics [61], [62]. For continuous transformations, Noether's theorem states that a physical system's invariance concerning a symmetry transformation corresponds to a conservation law. The symmetry of a system to shifts in space (time) gives rise to the conservation of momentum (energy), whose inner connection is revealed in the conservation of the 4-vector energy-momentum in relativistic theory. Another example is the conservation of angular momentum caused by rotational symmetry [61]. Two discrete transformations play a crucial role in the following discussion: time-reversal ($\hat{T}$) and space reversal or "parity" ($\hat{P}$). If a system obeys T-symmetry, its evolution can be time-reversed so that the system belongs to the same phase trajectory but going backward. In a T-symmetric electromagnetic system, wave transmission between two points in space does not change upon swapping source and receiver, giving rise to the reciprocity principle in Hermitian systems [63]–[72]. Incidentally, the correspondence between reciprocity and time-reversal symmetry always applies at the microscopic level because the equations of classical and quantum mechanics obey T-symmetry unless an axial dc bias, like a constant magnetic field, is applied. This is consistent with the principle of microscopic reversibility[73]–[75]. At the macroscopic level, the presence of loss can break T-symmetry, unless we also reverse the dissipation processes in time, converting them to gain [76]. The $\hat{T}$ operator flips the direction of time evolution, acting as complex conjugation. The laws governing classical mechanics, electrodynamics and quantum mechanics obey T-symmetry unless in the presence of a dc axial field [70], [71], time modulation [77], [78] or momentum biasing [79]. In turn, $\hat{P}$ plays an essential role in quantum mechanics (selection rules of atomic transitions), nonlinear optics (symmetry-forbidden harmonic generation) and condensed matter (spin-orbit coupling, valley polarization).

| Operator | Linear/antilinear | Action | Symmetry |
|---|---|---|---|
| $\hat{P}$ | linear unitary operator | $\hat{P}\psi(\mathbf{r},t) = \psi(-\mathbf{r},t)$ | $\hat{H}(\hat{\mathbf{p}},\hat{\mathbf{r}},t) = \hat{H}(-\hat{\mathbf{p}},-\hat{\mathbf{r}},t)$ |
| $\hat{T}$ | antilinear and antiunitary operator | $\hat{T}\psi(\mathbf{r},t) = \psi^*(\mathbf{r},-t)$ | $\hat{H}(\hat{\mathbf{p}},\hat{\mathbf{r}},t) = \hat{H}^*(-\hat{\mathbf{p}},\hat{\mathbf{r}},-t)$ |
| $\hat{P}\hat{T}$ | antilinear and antiunitary operator | $\hat{P}\hat{T}\psi(\mathbf{r},t) = \psi^*(-\mathbf{r},-t)$ | $\hat{H}(\hat{\mathbf{p}},\hat{\mathbf{r}},t) = \hat{H}^*(\hat{\mathbf{p}},-\hat{\mathbf{r}},-t)$ <br> Quant Mechanics: <br> $V(\mathbf{r}) = V^*(-\mathbf{r})$ <br> Optics: <br> $n'(\mathbf{r},\omega) = n'(-\mathbf{r},\omega)$ <br> $n''(\mathbf{r},\omega) = -n''(-\mathbf{r},\omega)$ |

**Table 1.** Discrete operators $\hat{P}$, $\hat{T}$ and $\hat{P}\hat{T}$ along with their action and symmetry requirements.

Let us consider now the properties of $\hat{P}$, $\hat{T}$ and $\hat{P}\hat{T}$, see Table 1. $\hat{P}$ changes the sign of all polar vectors [coordinates: $\mathbf{r} \to -\mathbf{r}$, momentum $\mathbf{p} \to -\mathbf{p}$, currents $\mathbf{j} \to -\mathbf{j}$, E-field: $\mathbf{E} \to -\mathbf{E}$, etc.], while axial vectors are not affected [magnetic field: $\mathbf{H} \to \mathbf{H}$, angular momentum: $\mathbf{L} \to \mathbf{L}$]. For example, for a scalar wave function $\psi(\mathbf{r},t)$ we have $\hat{P}\psi(\mathbf{r},t) = \psi(-\mathbf{r},t)$. The normalization of the wave function is preserved under spatial inversion, and therefore the parity operator is unitary, $\hat{P}^+\hat{P} = \hat{1}$, where $\hat{1}$ is the identity matrix. The action of $\hat{T}$ consists in replacing $t \to -t$, complex conjugation and changing the sign of all quantities that are defined via odd time derivatives [$\mathbf{p} \to -\mathbf{p}$], whereas time-independent physical quantities do not change $\mathbf{r} \to \mathbf{r}$. A scalar wave function $\psi(\mathbf{r},t)$ upon time-reversal transforms as $\hat{T}\psi(\mathbf{r},t) = \psi^*(\mathbf{r},-t)$. The action of the combined $\hat{P}\hat{T}$ operator consists in the subsequent action of $\hat{T}$ and $\hat{P}$ operators: $\hat{P}\hat{T}\psi(\mathbf{r},t) = \psi^*(-\mathbf{r},-t)$.

Wigner theorem [80] dictates that symmetry operators can be either linear and unitary, or antilinear and antiunitary. A linear operator does not change the complex factors, $\hat{O}_L[c\psi(\mathbf{r},t)] = c\hat{O}_L[\psi(\mathbf{r},t)]$, whereas antilinear operators leave them complex conjugated, $\hat{O}_{AL}[c\psi(\mathbf{r},t)] = c^*\hat{O}_{AL}[\psi(\mathbf{r},t)]$. Therefore (for $c = 1i$), $\hat{P}$ is a linear unitary operator, $\hat{T}$ is an antilinear and antiunitary operator and hence $\hat{P}\hat{T}$ is also antilinear and antiunitary. The transformation laws for operators can be derived from the fact that classical momentum and coordinates are defined by their mean values $\langle\hat{\mathbf{p}}\rangle$ and $\langle\mathbf{r}\rangle$, and have the same symmetry features. Hence, if $\hat{P}$ changes the sign of the polar vectors ($\mathbf{r}$, $\mathbf{p}$, $\mathbf{E}$) but leaves unaltered the sign of the axial vectors ($\hat{\mathbf{L}}$), the unitary transformation of the corresponding operators acts in the same way: $\hat{P}^+\hat{\mathbf{r}}\hat{P} = -\hat{\mathbf{r}}$, $\hat{P}^+\hat{\mathbf{p}}\hat{P} = -\hat{\mathbf{p}}$, $\hat{P}^+\hat{\mathbf{L}}\hat{P} = \hat{\mathbf{L}}$. Similarly, the action of the $\hat{T}$ operator changes the sign of the quantities defined through the first time derivative and hence the transformation of operators obey: $\hat{T}^+\hat{\mathbf{r}}\hat{T} = \hat{\mathbf{r}}$, $\hat{T}^+\hat{\mathbf{p}}\hat{T} = -\hat{\mathbf{p}}$, $\hat{T}^+\hat{\mathbf{L}}\hat{T} = -\hat{\mathbf{L}}$ [81].

All features of a physical system reside in the properties of the corresponding Hamiltonian. Namely, the response of a system to symmetry operations consists in the symmetry of the Hamiltonian with respect to the action of the corresponding operators. On the other hand, in quantum mechanics and optics, the fact that a system is symmetric to the action of a symmetry operation $\hat{O}$ means that the Hamiltonian and the operator commute, $[\hat{H}, \hat{O}] \equiv \hat{H}\hat{O} - \hat{O}\hat{H} = 0$. From the action rules of the operators $\hat{P}$, $\hat{T}$ and $\hat{P}\hat{T}$ follows that the Hamiltonian $\hat{H}(\hat{\mathbf{p}},\hat{\mathbf{r}},t)$ obeys P-symmetry if $\hat{H}(\hat{\mathbf{p}},\hat{\mathbf{r}},t) = \hat{H}(-\hat{\mathbf{p}},-\hat{\mathbf{r}},t)$, and T-symmetry if $\hat{H}(\hat{\mathbf{p}},\hat{\mathbf{r}},t) = \hat{H}^*(-\hat{\mathbf{p}},\hat{\mathbf{r}},-t)$. Their combination results in the requirement of a PT-symmetrical Hamiltonian, $\hat{H}(\hat{\mathbf{p}},\hat{\mathbf{r}},t) = \hat{H}^*(\hat{\mathbf{p}},-\hat{\mathbf{r}},-t)$, see Table 1. PT-symmetric Hamiltonians can support a real eigenvalue spectrum $\{\omega_k\}$ in their PT-symmetric phase when the corresponding eigensolutions $\{\psi_k\}$ satisfy PT-symmetry, i.e., $\hat{P}\hat{T}\psi(\mathbf{r},t) \to \psi^*(-\mathbf{r},-t) = \psi(\mathbf{r},t)$ and the commutation $[\hat{H}, \hat{P}\hat{T}] = 0$ is satisfied[2]. Under these conditions, although the energy is not necessarily conserved during the time evolution [82], the system has a real energy spectrum: given that we have balanced gain and loss, the system, on average, conserves energy in the symmetric phase, still remaining non-Hermitian. Interestingly, such a pseudo-Hermitian Hamiltonian can also support complex eigenvalues when it changes phase and $[\hat{H}, \hat{P}\hat{T}] \neq 0$, known as PT-symmetry broken phase, with non-PT-symmetric eigensolutions $\{\psi_k\}$: $\psi(\mathbf{r},t) \neq \psi^*(-\mathbf{r},-t)$. If $\{\omega_k\}$ are complex, then the Hamiltonian eigenfunctions are not eigenfunctions of the $\hat{P}\hat{T}$ operator [5]. Upon variation of a parameter, real eigenvalues are modified to complex ones through a second-order phase transition [83] associated with spontaneous PT-symmetry breaking.

Such pseudo-Hermitian Hamiltonians can be realized in various physical realms, but in electromagnetics they have become the subject of extensive research since the seminal works [10], [82]. Today, PT-symmetric systems hold a great promise for sensing [22], [84], imaging [85], lasing [86], cloaking and transformation optics[87], [88], new mechanisms of wave transport [89]–[91], topology [92], light scattering engineering [28], [93], [94], wireless power transfer [42], [57], [95], and for imaging [20], [96]–[99]. Before discussing these applications, it is important to highlight that the mentioned phase transition from symmetric to broken phase arises around a so-called exceptional point (EP). At an EP, the complex frequencies of two or more modes coalesce. Also the corresponding eigenstates become degenerate, so that the vector space loses one or more dimensions. In a PT-symmetric system, an EP can arise on the real frequency axis, and it can therefore be accessed with monochromatic signals. EPs can also be more generally found in the lower complex frequency plane in non-Hermitian systems. They differ from other degeneracy points, such as diabolic points (DPs) [100], at which two or more eigenvalues coalesce, but the eigenvectors remain distinct. DPs play an important role in chemistry [101], material science (Dirac points or Weyl nodes in semimetals) [102], [103] and are also exploited in optics and electromagnetics for sensing [104], [105], where they appear as 'mode splitting' points. For reasons discussed later, the sensitivity around EPs can exceed the one at DPs.

## 3 PT Symmetry in Quantum Mechanics and Optics

We start our discussion of $\hat{T}$ and $\hat{P}$ symmetry operation along with their combination, $\hat{P}\hat{T}$, by considering the single-particle Schrodinger equation

$$ i\hbar \frac{\partial \psi}{\partial t} = -\frac{\hbar^2}{2m} \frac{\partial^2 \psi}{\partial x^2} + V(x)\psi, \quad (1) $$

where $\psi$ is the wave function, $\hbar$ is the reduced Planck constant, and $m$ is the particle mass. The Hamiltonian $\hat{H} = -\frac{\hbar^2}{2m}\frac{\partial^2}{\partial x^2} + V(x) \equiv \hat{p}^2/2m + V(x)$ is composed of the kinetic energy operator $\hat{H}_k = -\frac{\hbar^2}{2m}\frac{\partial^2}{\partial x^2}$ and the potential energy operator $V(x)$. This Hamiltonian is invariant under a Hermitian conjugation, $\hat{H} = \hat{H}^\dagger$, if the potential is real-valued $V(x) = V^*(x)$, which ensures real eigen-energy and unitary time evolution. This theory describes conservative quantum systems, which are completely isolated from the surroundings and involving stable particles.

In the presence of open boundary conditions and possible energy fluxes in and out of the domain of interest, the potential $V(x)$ becomes complex, leading to non-Hermiticity. It is often commonly assumed that such an open system with complex potential $V(x) \neq V^*(x)$ inevitably results in a complex energy spectrum due to its non-conservative evolution. However, as discussed in the previous section, non-Hermitian systems can have a real-valued eigenspectrum if their Hamiltonian commutes with the $\hat{P}\hat{T}$ operator, i.e. $[\hat{H}, \hat{P}\hat{T}] \equiv \hat{H}(\hat{P}\hat{T}) - (\hat{P}\hat{T})\hat{H} = 0$. The Schrodinger equation (1) commutes with the $\hat{P}\hat{T}$ operator if and only if the potential energy satisfies $V(x) = V^*(-x)$. This requirement implies an even function for the real part of the potential ($\text{Re}[V(-x)] = \text{Re}[V(x)]$) and an odd function for its imaginary part ($\text{Im}[V(-x)] = -\text{Im}[V(x)]$).

To illustrate their ideas, Bender and Boettcher considered the class of PT-symmetric Hamiltonians $\hat{H} = \hat{p}^2 - (ix)^n$, where $n$ is real and describes the interaction strength of the particle with the potential. While this class of Hamiltonians is not symmetric under $\hat{P}$ or $\hat{T}$ separately, it is invariant under their combined operation. Even though the potential $V(x) = -(ix)^n$ satisfies the PT-symmetry condition $V(x) = V^*(-x)$, the spectrum of the Hamiltonian undergoes a PT-symmetry phase transition from the symmetric to broken phase as $n$ varies. Thus, the commutation relation $[\hat{H}, \hat{P}\hat{T}] = 0$ is a necessary but insufficient condition for the realness of the spectrum [4]. The physical reason is that the PT-symmetric phase is satisfied in a system with balanced net energy flux, and for values of $n$ corresponding to the PT-symmetry broken phase, this balance cannot be reached.

We are now ready to translate these concepts to the realm of electromagnetics. With this goal in mind, we examine an electromagnetic medium characterized by the refractive index $n(x) = \sqrt{\varepsilon(x)} > 0$, which is assumed to be complex $n = n' + in''$. If $n(x)$ varies slowly, this system can be described by the following equation for the slowly varying envelope of the transverse electric field $E$ with negligible transverse diffraction angles (a paraxial approximation of the wave equation) [11], [22]:

$$ i\frac{\partial E}{\partial z} + \frac{1}{2k_0}\frac{\partial^2 E}{\partial x^2} + k_0 n(x)E = 0, \quad (2) $$

where $k_0 = \omega/c$ is the wavenumber in free space and $\omega$ is the frequency. The link between Eqs. (1) and (2) can be established by substituting the time evolution by the spatial ($z$-dimension) propagation and introduce the complex potential $V(x) = k_0 n(x)$. As a result, the wave equation (2) can be represented in the form of (1) with an effective Hamiltonian

$$\hat{H} = -\frac{1}{2k_0}\frac{\partial^2}{\partial x^2} + k_0[n'(x) + in''(x)] \qquad (3)$$

whose eigenenergy represents the effective propagation constant (in contrast to the eigenfrequency) along the $z$-axis. We note that in this case, the action of $\hat{T}$ consists in flipping the propagation direction and complex conjugating.

From Eq.(3), we conclude that the Hamiltonian is PT-symmetric and can support a real spectrum if and only if $\mathrm{Re}[n(x)] \equiv n'(x) = n'(-x)$ and $\mathrm{Im}[n(x)] \equiv n''(x) = -n''(-x)$ are simultaneously satisfied [10], [82]. A trivial scenario corresponds to the case $n(x) = 0$. However, even in the presence of non-Hermitian $n''$, whose positive values correspond to absorption and negative values to material gain, achieved via stimulated emission in gainy materials or parametric phenomena [106], it is possible to have a real spectrum of eigenvalues provided that loss and gain distributions are balanced in space. This finding is not limited to the paraxial wave approximation and can be extended to full-wave Maxwell's equations, yielding, e.g., the requirement on the electric permittivity[88]:

$$\varepsilon'(\mathbf{r}, \omega) = \varepsilon'(-\mathbf{r}, \omega), \quad \varepsilon''(\mathbf{r}, \omega) = -\varepsilon''(-\mathbf{r}, \omega). \qquad (4)$$

Due to causality and its implications on the frequency dispersion of electromagnetic materials through Kramers-Kronig relations, Eq. (4) can be satisfied only at discrete frequencies [107], [108], implying that it is impossible to investigate a PT phase transition as frequency changes [109].

## 4 General properties

It is often convenient to describe PT-symmetric systems in terms of their S-matrix. This approach also allows us to discuss exotic scattering phenomena and the role that PT-symmetry and EPs have in these anomalies [28]. The scattering matrix $\hat{S}$ describes the coupling between incoming waves with amplitudes $\mathbf{i} = \{i_1, i_2, ...\}^\mathrm{T}$ and outgoing waves $\mathbf{o} = \{o_1, o_2, ...\}^\mathrm{T}$ within an arbitrary linear system [28], [110]

$$\begin{pmatrix} o_1 \\ o_2 \\ \vdots \\ o_N \end{pmatrix} = \begin{pmatrix} S_{11} & S_{12} & \cdots & S_{1N} \\ S_{21} & S_{22} & \cdots & S_{2N} \\ \vdots & \vdots & \vdots & \vdots \\ S_{N1} & S_{N2} & \cdots & S_{NN} \end{pmatrix} \begin{pmatrix} i_1 \\ i_2 \\ \vdots \\ i_N \end{pmatrix}, \qquad (5)$$

where $N$ is the number of channels. Usually, the amplitudes are normalized such that $|i_n|^2$ and $|o_n|^2$ are equal to the energy of the incoming and outgoing waves. In a two-port system, **Figure 1(a)**, the scattering matrix is $\hat{S} = \begin{pmatrix} r_{11} & t_{12} \\ t_{21} & r_{22} \end{pmatrix}$, where $r_{ii}$ and $t_{ij}$ stand for the reflection and transmission coefficients.

In Hermitian systems, the S-matrix eigenvalues $d_p$ are unimodular, $|d_p(\omega')|^2 = 1$, at any real frequency $\omega'$, as required by the unitarity of the scattering matrix, $\hat{S}\hat{S}^+ = \hat{I}$ [110]–[112]. In non-Hermitian lossy (gainy) systems, the scattering matrix is non-unitary at real frequencies, and the eigenvalues can be $|d_p(\omega')|^2 \leq 1$ ($|d_p(\omega')|^2 \geq 1$). In the complex frequency plane, $\omega = \omega' + i\omega''$, the scattering eigenvalues $|d_p(\omega)|$ are generally not bounded. We can found point-like singularities $|d_p(\omega)| = \infty$ (scattering pole) or $|d_p(\omega)| = 0$ (scattering zero). The poles correspond to self-sustained fields (eigenmodes) [113], whereas the zeros are scattering-less states (perfect absorption or capturing). It can be rigorously proven that the knowledge of these special points in the complex plane allows a complete description of the electromagnetic features of the linear system under consideration at all frequencies [28], [114]–[116]. Similarly, engineering these peculiarities in the complex frequency plane allows tailoring structures with exotic scattering phenomena of choice [28].

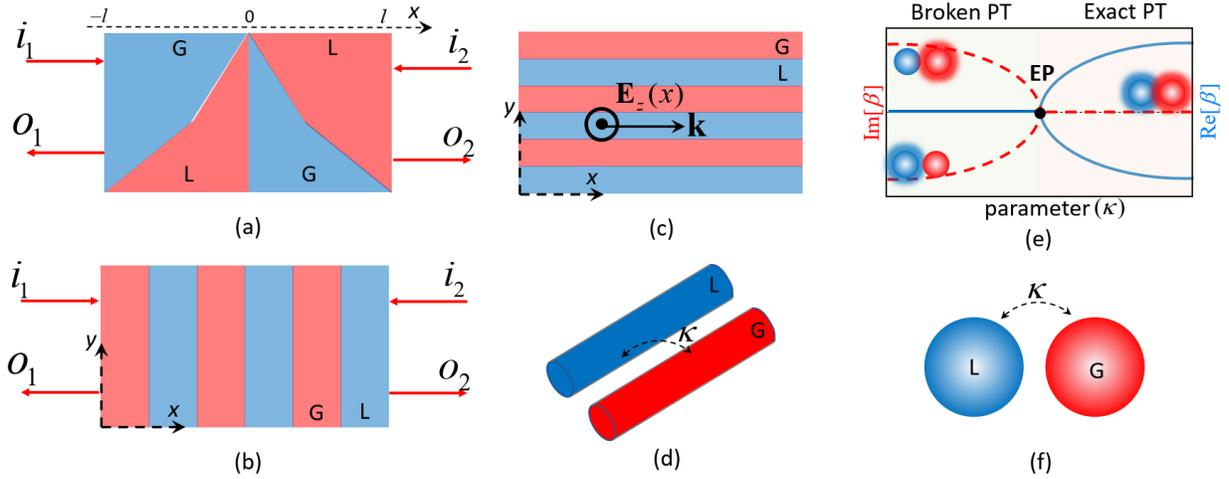

**Figure 1.** (a) S-matrix description of a linear system with gain (G) and loss (L). (b) Multilayered two-port structure. (c) Geometry of an analytically solvable problem. (d,f) System of two coupled waveguides (d) or localized resonators (f) coupled with the coupling strength $\kappa$. (e) Schematic transition in the eigenvalues from purely real (exact PT symmetry) to purely imaginary (broken PT symmetry) in an active balanced system. EP stands for exceptional point.

The S-matrix is a powerful approach to investigate the scattering from systems with well-defined boundaries and incoming and outgoing channels, e.g., the multilayered structures [19], [117]–[119] shown in Figure 1(b). The S-matrix of a generalized PT-symmetric two-port system is transformed as (see Appendix A)

$$(\hat{P}\hat{T})S_{nm}(\hat{P}\hat{T}) = S_{nm}^{-1}. \tag{6}$$

Let us now define a scattering condition under which a non-Hermitian two-port system, as in Figure 1(a), is in the exact PT-symmetric phase. Under T transformation, the propagation direction is flipped, and the amplitudes are conjugated, $\{i_n\} \xrightarrow{\hat{T}} \{o_n^*\}$. Since $\hat{P}$ changes the propagation direction and $\hat{T}$ interchanges the waves propagating from left to right and from right to left, the $\hat{P}\hat{T}$ operator results in $i_1 = o_2^*$ and $i_2 = o_1^*$. Since the S-matrix connects the amplitudes of the incident and scattered waves, they should have pairwise equal moduli, $|i_1| = |o_1|$ and $|i_2| = |o_2|$. Combining these two results, we find that the amplitudes from the left and right must be equal in the PT-symmetric phase, $|i_1|=|i_2|$ and $|o_2|=|o_2|$. Hence, if we assume $i_1=1$ ($o_2=1$), the amplitude of the second incident wave can be written as $i_2=\exp(i\varphi)$ [$o_1=\exp(-i\varphi)$], where $\varphi$ is an arbitrary phase. The scattered field in the left (right) port is the sum of the transmitted wave from the right (left) and the reflected wave, giving rise to $\exp(-i\varphi)=t\exp(i\varphi)+r_l$, and $1=r_r\exp(i\varphi)+t$. Combining these two expressions, we obtain $(r_l-r_r)/t=-2i\sin(\varphi)$, which gives a condition for the PT-symmetric phase:

$$\left|\frac{r_l-r_r}{t}\right|\leq 2. \tag{7}$$

If Eq. (7) is satisfied, the system under consideration is in its PT-symmetric phase; otherwise, it is in the broken phase[120]. Another test for PT-symmetry can be introduced using energy considerations: in the exact PT-symmetric case, both eigenstates act as pseudo-Hermitian, and a generalized form of power conservation holds: $P=T+\sqrt{R_rR_l}=1$, where $T=|t|^2$ and $R_{r,l}=|r_{r,l}|^2$ are the transmission and reflection coefficients. In the broken phase, $P>1$ or $P<1$ for the gainy and lossy eigenstates, respectively [121].

An example of an analytically solvable PT-symmetric system is presented in Figure 1(c). We consider a linearly polarized wave with $z$-polarized electric field $E_z(y)$ propagating in such a layered gain-loss structure. This 2D scalar problem is described by the Helmholtz equation [22], [109], [122] $\Delta_{x,y}E_z(x,y)+k_0^2\varepsilon(y)E_z(x,y)=0$, where $\Delta_{x,y}\equiv\partial^2/\partial^2x+\partial^2/\partial^2y$. Since the permittivity does not depend on the $z$-coordinate, we can apply variable separation $E_z(x,y)=f(y)g(x)$ to solve the Helmholtz equation, leading to the equations $\partial^2f(y)/\partial y^2+k_0^2\varepsilon(y)f(y)=k_x^2f(y)$ and $\partial^2g(x)/\partial x^2+k_x^2g(x)=0$. The solution of the latter one is a sum of two counterpropagating waves $g(x)=A_1\exp(ik_xx)+A_2\exp(-ik_xx)$. The former equation coincides with the 1D stationary Schrodinger equation with Hamiltonian $\hat{H}=\hat{p}^2/2m+V(x)$, with replacement $p^2/(2m)\rightarrow\partial^2/\partial y^2$, $V(x)\rightarrow k_0^2\varepsilon(y)$, $E\rightarrow k_x^2$, and $\psi(x)\rightarrow f(y)$. In this case, the operation $\hat{P}$ consists in inversion along the $y$-axis, and $\hat{T}$ leads to complex conjugation.

## 5 Exceptional points

In the S-matrix approach, EPs arise at the coalescence of two poles. In open systems, they can occur in the complex frequency plane in passive systems or on the real frequency axis in PT-symmetric systems. The latter case is of most interest, since it enables access to EPs for monochromatic excitations. The emergence of EPs at real frequencies is possible in systems with gain, and it requires at least two modes capable of strong coupling [22], [29]. In fact, this type of EPs appears exactly at the boundary between the weak and strong coupling regimes[123]. It is worth stressing as a side note that EPs are ubiquitous [124], they go far beyond PT-symmetric systems and can be found in various phenomena whenever the spectrum depends on at least two parameters ($\alpha$, $\beta$) as $\sim\sqrt{\alpha^2-\beta^2}$, where a phase transition occurs at $\alpha=\beta$ and the spectrum goes from being real to complex-valued. For example, an EP manifests itself as a critical point between damped and overdamped dynamics of an oscillator or a pendulum, vanishing group velocity in gratings [125], total internal reflection at the interface of two materials [126], the cut-off frequency of a closed waveguide[127]. We do not touch these aspects in this paper and restrict ourselves to EPs emerging in PT-symmetric systems, which are particularly interesting for their associated features.

As an illustration of EPs in PT-symmetric systems, we consider a pair of coupled single-mode waveguides, where balanced gain and loss are implemented in the two waveguides, respectively, Figure 1(d) [24], [128]. This problem is equivalent to two coupled resonators, Figure 1(f). The equation governing these systems reads $\frac{d}{d\xi}\begin{pmatrix}a_1\\a_2\end{pmatrix} = -i\hat{H}\begin{pmatrix}a_1\\a_2\end{pmatrix}$, where $a_1$ and $a_2$ are either waveguiding modes [Figure 1(d)] or localized resonator modes [Figure 1(f)], $\xi$ is either the coordinate or time, respectively. The Hamiltonian of these two-mode systems can be derived from the standard coupled-mode equations [43], [57], [59], [129]–[136]

$$\hat{H} = \begin{pmatrix} \beta_{0,1} - i\gamma_1 & \kappa \\ \kappa^* & \beta_{0,2} - i\gamma_2 \end{pmatrix}, \tag{8}$$

where $\beta_{0,1}$ and $\beta_{0,2}$ are either propagation constants in each waveguide or eigenfrequencies of the localized modes, $\gamma_i$ takes account of the gain/loss magnitude in the waveguide (mode) $i$, and $\kappa$ denotes the reciprocal coupling between the waveguide pair. Solving the eigenstate problem gives the eigenvalues of (8)

$$\beta_\pm = \beta_{0,\mathrm{ave}} - i\gamma_{\mathrm{ave}} \pm \sqrt{|\kappa|^2 + (\beta_{0,\mathrm{dif}} + i\gamma_{\mathrm{dif}})^2}, \tag{9}$$

where $\beta_{0,\mathrm{ave}} = (\beta_{0,1} + \beta_{0,2})/2$, $\beta_{0,\mathrm{dif}} = (\beta_{0,1} - \beta_{0,2})/2$, $\gamma_{\mathrm{ave}} = (\gamma_1 + \gamma_2)/2$, $\gamma_{\mathrm{dif}} = (\gamma_1 - \gamma_2)/2$. In the particular case when $\beta_{0,1} = \beta_{0,2} \equiv \beta_0$ and $\gamma_2 = -\gamma_1 \equiv \gamma$, the eigenvalues of the two supermodes become $\beta_\pm = \beta_0 \pm \sqrt{|\kappa|^2 - \gamma^2}$. Eq. (9) shows that, if the coupling is less than a certain critical value ($\kappa < \kappa_{\mathrm{PT}} = \gamma$), the system possesses two modes, one lossy and one amplifying. When excited, the lossy mode decays exponentially in time, whereas the gainy one exhibits exponential growth. However, if the coupling strength is large enough ($\kappa > \kappa_{\mathrm{PT}}$), the system resides in the strong coupling regime for which the coherent energy exchange between the elements compensates for the decay and stabilizes the system at the real frequency axis, Figure 1(e). The critical point $\kappa = \kappa_{\mathrm{PT}}$ gives rise to an exceptional point (EP) [11], [28], [82], [84], [86], [137], [138]. Exactly at the transition threshold where $\gamma_2 = -\gamma_1$, the two supermodes coalesce to a single eigenstate - $(1,i)^{\mathrm{T}}/\sqrt{2}$ with propagation constant $\beta_0$, featuring a non-Hermitian degeneracy. Despite its simplicity, this model (8) for a non-Hermitian dimer pair has become a powerful tool to describe PT-symmetric responses in various transversely coupled systems, such as coupled microlasers [43], [139], [140], silica microtoroids [52], [141] and photonic crystal cavities [142].

PT-symmetry requires a balance of loss and gain, which is challenging to achieve from an experimental perspective, especially as the frequency of operation grows. Nevertheless, some of the interesting physics of PT-symmetry can also be explored in non-Hermitian systems with a loss offset, and loss-only structures. In fact, the Hamiltonian $\begin{pmatrix} \beta_0 - i\gamma_1 & \kappa \\ \kappa^* & \beta_0 - i\gamma_2 \end{pmatrix}$ of a lossy two-mode system can be represented as the sum of PT-symmetric ($\hat{H}_{PT}$) and decaying ($\hat{H}_L$) subsystems, introducing the following notations $\gamma_+ = \gamma_1 + \gamma_2$ and $\gamma_- = \gamma_2 - \gamma_1$:

$$\hat{H} = \hat{H}_L + \hat{H}_{PT} = \begin{pmatrix} -\gamma_+/2 & 0 \\ 0 & -\gamma_+/2 \end{pmatrix} + \begin{pmatrix} \beta_0 + i\gamma_-/2 & \kappa \\ \kappa^* & \beta_0 - i\gamma_-/2 \end{pmatrix}, \tag{10}$$

where $\hat{H}_{PT}$ duplicates the PT-symmetric Hamiltonian (10) for balanced gain and loss rates. Thus, the effects of PT-symmetry and EPs can be observed in passive systems with a loss offset. This approach simplifies the technical challenges of realizing active PT-symmetric systems[24], [28], [143]–[145]. The EP associated with such passive PT-symmetry resides in the lower complex half-plane, and it can be observed with either parameter variations or complex excitations [26]. In an experiment with loss-only structures, a phase transition manifests itself as a change in the mode symmetry from asymmetric in the PT-symmetry broken to symmetric in the PT-symmetric phase as the material loss is reduced [24]. Lossy and gain-loss PT-symmetry and EPs are anyhow strictly different. For example, in only lossy systems lasing or lasing-CPA states that emerge in PT-symmetric systems cannot be achieved.

In Eqs. (8) and (10), the coefficient $\kappa$ is assumed to be a real parameter, which corresponds to a near-field (quasistatic) in-phase coupling. In principle, however, the coupling can be complex, involving, e.g., radiation modes or retardation effects [146].

The two-mode problem represents the simplest toy model embracing the key features of PT-symmetry and EPs. In general, the coalescence of more than two eigenmodes creates a higher-order exceptional point [147]–[151]. A higher-order exceptional point, e.g., is formed when three or more eigenvalues simultaneously coalesce. The coalescence of multiple eigenvalues requires a delicate variation of parameters within a larger parameter space and is much more challenging to achieve [150], [152]. However, as a reward, this gives even more interesting features, e.g., for sensing purposes. In the next section, we discuss the place that EPs occupy among other light scattering phenomena.

## 5 Classification of scattering anomalies

The S-matrix approach discussed in the previous section allows us to classify the scattering phenomena emerging in various scenarios of interest. Several anomalies in light scattering are sketched in Figure 2 in terms of the positions of poles and zeros of the S-matrix. We begin the discussion with a Hermitian system with "weak ports", i.e., when the poles and zeros lie very close to the real axis. Increasing radiation leads to a repulsion of poles and zeros to the complex plane. As discussed above, since we are neglecting material loss, the poles and zeros are connected via complex conjugation and are located symmetrically with respect to the real axis, Figure 2(a) [28], [106], [153], [154].

An interesting phenomenon can arise in this lossless scenario when a pole and the corresponding zero get very close to one another on the real axis. The pole nearby the real axis supports an unboundedly large Q-factor, which grows the closer the pole gets to the real axis. This is possible due to the presence of the zero at the same real frequency. This effect is called a bound state in the continuum (BIC), or embedded eigenstate (EE) [155]–[158], Figure 2(b). BICs can be realized in either symmetry-protected scenarios, e.g., in the case of a hedgehog-like collection of dipoles longitudinally arranged over a sphere or a periodic 2D array of in-phase dipoles, or exploiting the destructive interference of at least two strongly coupled resonant modes coupled to the same radiation channel[155], [159]–[166]. Due to their very high Q-factor and topological features, BICs offer exciting perspectives for applications in sensing[167], lasing[168]–[170], nonreciprocity when combined with nonlinearities[171], thermal emission[172], [173] and energy harvesting [174].

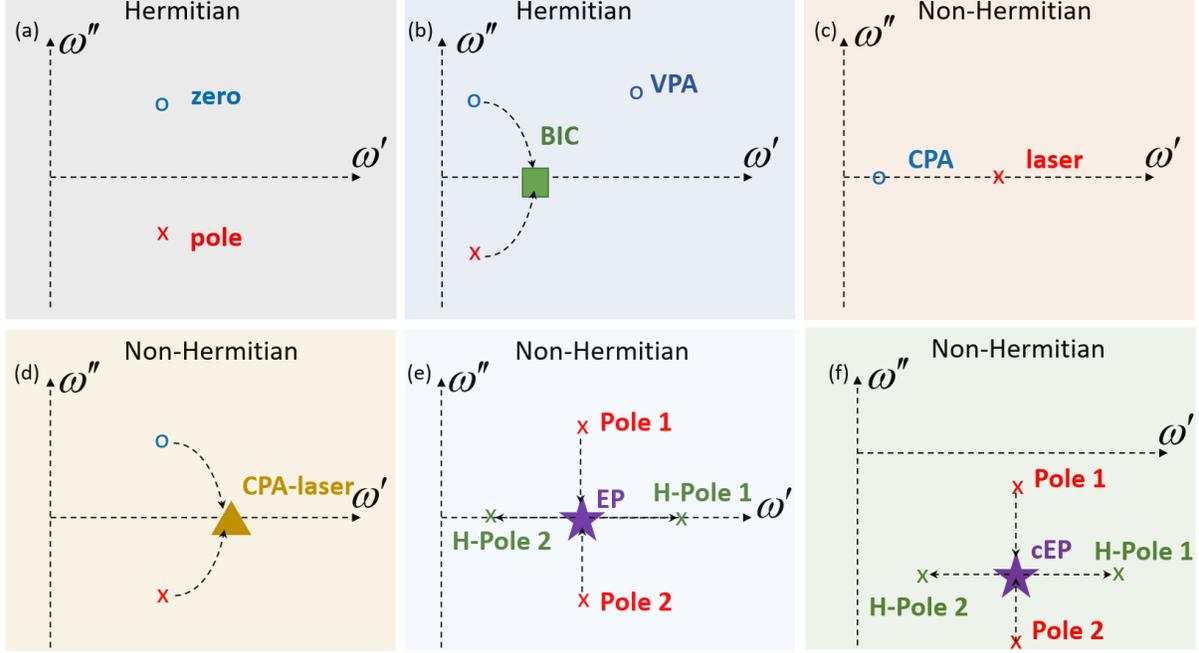

**Figure 2.** Anomalies in light scattering in terms of the S-matrix poles and zeros. (a) In a system with neither material nor radiative loss, poles and zeros lie unboundedly close to the real axis. Adding radiation loss repulses the poles and zeros to the complex plane, where they are connected via complex conjugation, i.e., they are located symmetrically to the real axis. (b) Tailoring a system symmetry or utilizing ENZ materials makes it feasible to coalesce a pole and a zero at one frequency point, giving rise to a BIC. In the absence of material loss, a system can have a zero in the upper half-plane, which can be excited by a complex wave (VPA). (c) Adding loss (gain), we can move a zero (pole) to the real frequency axis, giving rise to perfect absorption (lasing). In a general case of multiple ports, the perfect absorption can be generalized to CPA. (d) In an active system, the coalescence of zeros and poles does not destroy the points but instead leads to the coexistence of CPA and lasing effects at the same frequency. (e) In an active system, a variation of the coupling interaction strength between at least two modes (gainy pole 1 and lossy pole 2) can result in their coalescence on the real axis in the balanced case, giving rise to an EP. (f) The coalescence of two poles can also occur in the lower complex half-plane, giving rise to a cEP. Further increases of the interaction parameter lead to strong coupling and the formation of two hybrid modes (H-Pole 1 and H-Pole 2), also known as dressed states.

Adding loss (gain), we can move a zero (pole) to the real axis, giving rise to perfect absorption (lasing), Figure 2(c). In the general case of multiple ports, the perfect absorption can be generalized to coherent perfect absorption (CPA) [28], [175]–[178]. CPA and lasing can be considered the time-reversal ($\hat{T}$) realizations of one another. The CPA enables a device that completely absorbs coherent inputs with specific relative amplitudes and phases, corresponding to the vector associated with the scattering zero, enabled by destructive interference between the transmission and reflection outside the system and their constructive interference inside the cavity. The overall absorption and transmission of the system can be dynamically controlled by carefully tuning the relative phase between the signals at the various ports [28], [175].

Exploiting coherence to control optical phenomena offers rich opportunities beyond enhancing absorption. It has been recently shown that instead of adding material loss to a system to achieve CPA, we can exploit radiation losses, based on which coherent excitation leads to manipulation of radiation [179]. Another example is coherently enhanced wireless power transfer [180], wherein coherent excitations were exploited for the enhancement and control of energy transfer. It is also possible to induce CPA-like responses in lossless systems by engaging the S-matrix zeros in the complex frequency plane, Figure 2(b). In this case, we need to tailor the incident fields in time such that their temporal profile matches the exponentially growing fields associated with the complex frequency of the zero. During this transient, the scattering from the structure vanishes, as if the structure were perfectly absorbing, even though it does not support loss or absorption, giving rise to virtual perfect absorption (VPA). For these complex frequency excitations, the system does not reach a steady-state, and can continue to store the impinging electromagnetic energy, rather than converting it into other forms of energy as in the case of conventional absorbing systems. Hence, VPA can be used for efficient storage and release in optical memories [181], [182]. VPA can also be extended to general regimes in which the level of material loss is nonzero, and part of the energy is absorbed, part stored in the system [174].

Lasers, i.e., coherent sources and amplifiers, are active optical devices for which one S-matrix pole reaches the real axis [106], Figure 2(c). In order to achieve lasing, the system has to contain material gain and a high-Q cavity, providing the required feedback. However, a high-Q cavity, such as a whispering gallery, dielectric microdisks, defected photonic crystals etc., has spatially or spectrally degenerate modes (poles) that compete for the limited resource of optical gain. This results in the emergence of several active modes and overall degradation of monochromaticity and photon coherence of the laser. Remarkably, it has been recently demonstrated that PT-symmetry can turn such a multimode laser into a single-mode laser [43], [140]. The approach consists in designing a laser consisting of two resonators that are equivalent except that one is active, but another one is lossy. Pumping the active resonator results in a PT-symmetry regime for all degenerate modes except for two, entering the PT-symmetry broken phase, with one of them being lossy and another one gainy. The gainy mode of the PT-symmetry broken configuration manifests itself as a laser. Since the laser emission is not coupled to other modes, the lasing mode has good threshold characteristics. The maximum achievable gain, defined as the contrast between the gain $g_0$ at the lasing mode and the gain $g_1$ of the next competing mode, is controlled by the EP[43]

$$g_{max}^{PT} = g_{max}\sqrt{\frac{g_0/g_1+1}{g_0/g_1-1}}, \qquad (11)$$

where $g_{max} = g_0 - g_1$ is the maximum achievable gain contrast for a laser without PT-symmetry. This square-root behavior is a consequence of the topology in the vicinity of an EP in non-Hermitian systems. These concepts have been applied to various laser configurations and platforms. EPs and associated transitions to single-mode lasing were observed in photonic crystal lasers [142] and in electrically pumped and coherently coupled vertical-cavity surface-emitting laser (VCSEL) arrays [183].

In a non-Hermitian active system, the coalescence of an S-matrix zero and pole does not destroy the points but instead leads to the coexistence of CPA and lasing at the same frequency for different forms of excitations. This effect has been theoretically proposed in [120], [184], where it was shown that this condition can emerge only in the PT-broken phase. Indeed, only in this regime the zero (perfect absorption) and pole (lasing) states coalesce after anti-crossing as the balanced gain/loss is increased. In contrast to an EP state, the eigenstates of the S-matrix do not coincide in the CPA-laser state, and hence the excitations of the zero and pole undergo different relative phases at the system ports [106], [185], [186]. In [187], a CPA-laser was implemented using an active III-V semiconductor waveguide modulated by an on-top deposited loss Cr/Ge grating. The measured lasing and absorption spectra validated the claim of efficient light control at the operation wavelength, showing that, by controlling the relative phase offset of the input signals, either coherent amplification ('ON' state) or absorption ('OFF' state) were enabled within a single device.

By varying the coupling interaction strengths between two modes (gainy pole 1 and lossy pole 2) of an active system can result in their coalescence on the real frequency axis, giving rise to an EP. The EP in a PT-symmetric structure coincides with the spontaneous symmetry-breaking threshold, at which the unbroken PT-symmetric phase abruptly transits to the PT-broken phase. EPs are of great interest for various applications. A laser system can oscillate between the PT unbroken and broken phases when operated in the EP proximity, at which a loss-induced revival of lasing counterintuitively arises [188], [189]. The coalescence of eigenstates at the EP also enables unidirectional light transport, leading to the observation of anisotropic transmission resonances [51], [53], [190], [191] and directional power flow in microlaser cavities [192]–[194]. Moreover, the accidental coalescence of eigenmodes offers rich topological features, enabling dynamic chiral-mode conversion [59], [195], [196], and the enhancement of sensitivity for optical sensing [32], [37], [197].

A significant challenge for practical wireless power transfer (WPT) systems is to achieve stable power transfer with high and robust transfer efficiency upon variations of the coupling conditions. Recently, this issue has been addressed with systems combining nonlinearity with PT-symmetry and EPs [57], [58], [198], [199]. In [200], it was demonstrated that this approach can be applied for drone-in-flight wireless charging. The reported experimental results have shown that when the flying drone hovers in a confined three-dimensional volume of space above the WPT system, the stable output power is maintained with a high transfer efficiency of 93.6%.

PT-symmetric phenomena become even more exciting when nonlinear phenomena are considered [13], [201], [202] and particularly in the case of optical solitons [203], [204], as experimentally investigated in time-domain lattices [205]. In nonlinear systems, we cannot generally employ our S-matrix description, and other unusual phenomena can emerge in combination with the described phase transitions, such as multi-stable operation and large asymmetries between the rise and decay times of the resonant states [171].

Finalizing the discussion of the scattering phenomena captured in Figure (2), we note that the coalescence of two poles can also arise in the lower complex half plane, giving rise to complex EPs (cEPs). In the strong coupling regime, we can observe the emergence of hybrid modes (H-Pole 1 and H-Pole 2), or dressed states, Figure 2(f). Even though such cEPs reside in the complex frequency plane, they can be revealed with parameter variations [23] or complex frequency excitations [26]. In contrast to VPA, cEPs are excited with exponentially decaying signals because cEPs lie in the lower half-plane ($-|\omega''|$).

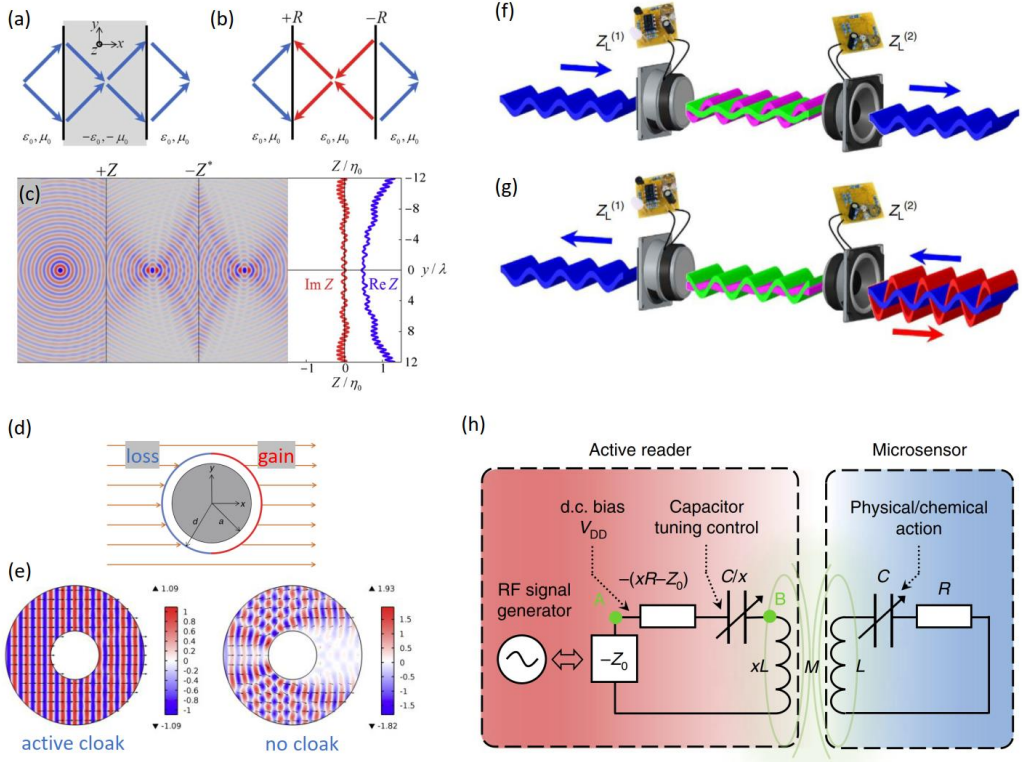

**Figure 3.** Applications of PT-symmetric systems. (a) Conventional negative refraction in a passive double-negative medium for light rays emitted by a source placed on the slab's left side. The power flows away from the source, and the phase velocity in the slab is backward. (b) Negative refraction using PT-symmetric metasurfaces with real surface impedances R and -R. In this active scenario, both power flow and the phase velocity are directed from the active surface to the passive one, and negative refraction is obtained without the need for a bulk metamaterial. (c) Planar focusing using a pair of PT-symmetric metasurfaces for a line current source placed on the left of the structure. An inhomogeneous surface impedance focuses all propagating spatial harmonics[98]. (d) PT-symmetric cloak[206]. The left and right portions of the surface (blue and red in the figure) have loss and gain, respectively. The lossy part is tailored to absorb all of the impinging power; the gain part emits the same amount of power, synchronized to the impinging signal. (e) Total electric field (color plot, a snapshot in time, normalized to the incident amplitude) and power (vectorial plot) distribution for a PEC cylinder with a $a = 2\lambda$ with (left) and without (right) active cloak. (f), (g) A parity-time invisible acoustic sensor[20]. A PT-symmetric acoustic system is realized using a pair of electromechanical resonators (loudspeakers) loaded with properly tailored electrical circuits. The left loudspeaker is operated as a sensor by loading it with an absorptive circuit, while the other forms an acoustic gain element. Their combination is a compact PT-symmetric unit cell that is transparent from the left (f), while it can simultaneously extract the impinging signal. On the contrary, the system is highly reflective when excited from the right (g). (h) Generalized parity-time symmetry (PTX) symmetric telemetric sensor system, where $x$ is the scaling coefficient of the reciprocal-scaling operation X. If $x = 1$, the PTX system converges to the PT-symmetric case. In the closed-loop normal mode analysis, an RF signal generator with a source impedance $Z_0$, connected to the reader is represented by $-Z_0$[207].

# 6 Applications of PT-symmetric structures

In this section we discuss interesting electromagnetic wave phenomena arising in PT-symmetric structures, which may be of special interest to the antennas and propagation community. To start, negative refraction and focusing have been a popular topic in the metamaterials community at large. It is well known that optical structures, such as lenses, prisms, etc., suffer from aberrations due to their curvature, and only a flat mirror could form an ideal imaging device. However, Veselago [208] and Pendry [209] excited the physics community at large showing that negative refractive index materials could enable flat lenses forming an aberration-free image, Figure 3(a). Although negative refractive index has been shown in various artificial electromagnetic materials or metamaterials [209]–[212], it has been soon realized that their unusual imaging effects are hindered by the presence of loss and nonlocalities in realistic implementations. Interestingly, PT-symmetry offers an alternative route to negative refraction and planar focusing that can go beyond these limitations [96], [98]. A pair of PT-symmetric thin metasurfaces with balanced gain and loss, i.e., with surface impedances satisfying $Z_{\text{left}} = -Z_{\text{right}}^*$, Figure 3(b), can be made reflectionless (i.e., impedance-matched). At the same time, in steady-state the gainy portion of the bilayer emits two phase conjugated replicas of the impinging signal, synchronized to it, one traveling backward and absorbed by the lossy surface, the other one traveling at the output port and forming an aberration-free image on the other side of the bilayer. The structure essentially replicates the response of a negative index slab of thickness equal to the distance between the PT-symmetric surfaces.

If their surface impedance is local, however, the impedance matching condition cannot be achieved for all incidence angles, implying that a focusing effect for a wide-angle image can be achieved only for inhomogeneous impedance distributions, as in Figure 3(c). However, in this operation aberrations may arise, since the system is designed for a specific focal point, as in a regular lens. This challenge can be overcome by synthesizing nonlocal impedance responses, for instance using PT-symmetric multilayered planar slabs [96]. In this case, we obtain a response identical to an ideal impedance-matched negative-index slab, but with the advantage of not having to use a large array of resonant elements, prone to loss and imperfections. Here loss actually are leveraged for the phenomenon, by properly balancing them with gain. The result is an aberration-free planal lens, implementing the imaging properties of planar absolute optical instruments. These concepts can be implemented with optically or electrically pumped materials [106] or with metasurfaces loaded with complementary positive and negative impedance elements.

A 1-D implementation of this PT-symmetric metasurface pair has been recently implemented in acoustics, Figure 3(f,g), using a pair of loudspeakers (electromechanical resonators) loaded with tailored electrical circuits, one implementing the time-reversed response of the other one. In this configuration, in addition to negative-index propagation, the system can support the unique sensing functionalities [20]. The left resonator absorbs both the incident signal and the conjugated signal fed by the active resonator and impinging from the opposite side, operating as a very efficient sensor. At the same time, the gain element eliminates any shadow, realizing a unique sensor with strong absorption and zero scattering, Fig. 3(f). The system is non-Hermitian, therefore does not have to comply with the power conservation rules of lossless systems. Indeed, when excited from the gainy side, while still being transparent because of reciprocity, the PT-symmetric pair is at the same time highly reflective, Fig. 3(g). Therefore, this unusual sensor implements an anisotropic transmission resonance, with a functionality of great interest for advanced metrology [226].

Camouflaging and invisibility represent another application of great interest in the field of metamaterials [28], [213]–[219]. Similar challenges arise, the loss and bandwidth limitations of passive metamaterials inherently limit the applicability of these concepts. Fundamental studies on the causality of the response of passive metamaterials have shown that any passive cloak is fundamentally limited in terms of its bandwidth of operation [28], [220], and only active implementations can realize broadband scattering suppression [221]-[223].

PT-symmetric metasurfaces with balanced combinations of gain and loss have been recently shown to be capable of achieving invisibility phenomena beyond the limitations of passive metamaterials. Similar to the previous demonstrations, the approach consists in perfect absorption of the impinging energy on one side of the cloaked object with simultaneous reradiation of the impinging signal on the other side. This approach has been used to realize unidirectional electromagnetic cloaks for large objects [206], as shown in Figure 3(d). Figure 3(e) shows the total electric field and power distribution for a PEC cylinder ($a = 2\lambda$) with (left) and without (right) active cloak. We note that, in contrast to other types of active cloaks that typically exploit field sensors that detect the external wave and assist in the generation of the auxiliary coherent wave radiated by an array of antennas in order to cancel the scattered field [223]–[225], the PT-symmetric cloak automatically synchronizes itself with the impinging signal because of its nature, without the need for any external control.

Figure 3(h) shows another layout for a PT-symmetric sensor for radio/microwave telemetry, exploiting inductive coupling between two coils connected to lossy and gainy parts, respectively. Parameter $x$ in the figure provides an additional degree of freedom to control the system. If $x = 1$, -R and R in both shoulders are balanced, and the system is tuned in the PT-symmetric configuration with an EP response. This type of sensor has been widely used across the spectrum, from radio-frequencies to optics [32], [227], [228], as we discuss in the following section in more details. In [207], a generalized PT-symmetry condition (PTX-symmetry) for enhanced sensing has been introduced, using the tunability available in circuit elements. This approach enables miniaturized wireless microsensors with high spectral resolution and high sensitivity.

In all these applications, it is important to carefully consider the stability of these designs, since they involve active elements they may lead to oscillations unless suitably designed. Dispersion engineering of the gain profile may be necessary to ensure an absolutely stable operation [97].

# 7 Sensing at an EP and the influence of noise

We have already mentioned how EPs in PT-symmetric systems can be beneficial for sensing applications. Theoretically, we can achieve extremely high sensitivities using EPs, since the eigenvalues around these special points have diverging susceptibility over a small parameter change $\epsilon$, if the eigenvalue difference is $\delta\beta \sim \epsilon^{1/n}$, then $d(\delta\beta)/d\epsilon = \epsilon^{1/n}/\epsilon$, where $n$ is the order of EP [37], [229], [230]. For instance, in the system in Fig. 1, corresponding to a second-order EP, the evolution of the eigenvalues around the EP has a square-root behavior, as evident in Fig. 1e. This super-linear response is superior to any conventional sensor, even those operating at a DP, which are all characterized by a linear variation of the eigenvalues for small perturbations. However, there is still an open discussion in the literature on how the necessary presence of noise in any practical system may affect this sensitivity. Generally, noise emerges due to the fluctuation-dissipation theorem, which connects dissipation or gain in a system with the spectral density of noise [231] and it is applicable to classical and quantum systems. Specifically, in quantum systems, the noise influence can be most significant. Thus, studying the system response when noise is considered and finding techniques to reduce the negative influence of noise become important directions for investigation.

There is a straightforward way to introduce noise in the basic model of two coupled modes: we can add noise in the right-hand side of the coupled-mode equations

$$\frac{d}{dt}\begin{pmatrix} a_1 \\ a_2 \end{pmatrix} = \begin{pmatrix} \beta_{0,1} - i\gamma_1(\eta_1) & \kappa \\ \kappa^* & \beta_{0,2} - i\gamma_2(\eta_2) \end{pmatrix} \begin{pmatrix} a_1 \\ a_2 \end{pmatrix} + \begin{pmatrix} \sqrt{\eta_1} F_1(t) \\ \sqrt{\eta_2} F_2(t) \end{pmatrix}, \tag{12}$$

where $F_i(t)$ describes stochastic fluctuations. The spectral density of noise $\eta_i$ can be obtained from observing the full system and its reservoirs [232], [233], or can be added phenomenologically.

If we consider the same classical system as described in (12) and take into account the fluctuations of the uncoupled resonator frequencies, we find that the noise modulates the EP conditions [132]. Statistical averaging worsens the spectral features, and the spectrum near the EP broadens. These phenomena significantly limit the sensitivity in the presence of noise, and the temporal fluctuations in the detuning and gain of the system lead to at least a quadratic growth of the optical power in time, implying that maintaining operation at the EP over a long period can be challenging.

More radical conclusions about sensing near EPs have been made: by analyzing a pair of coupled resonators, in [234] fundamental bounds on the output signal power and signal-noise ratio have been derived, indicating that gain-loss PT-symmetric systems may not be necessarily beneficial for sensing applications. An all-loss pair of resonators operated near its EP, despite the smaller output signals, can achieve equal or better measurement rates compared to a system with balanced loss and gain near an EP when noise is considered. This originates from the fact that gain amplifies not only the signal but also the noise. In the presence of noise, nonreciprocity can be a powerful tool for sensing, and nonreciprocal devices can overcome the fundamental bounds found for reciprocal sensors. Considering quantum Fisher information, we can calculate the signal-to-noise ratio (SNR) for EP sensors [230], [235], confirming the conclusion of [234]. The discussion is still open, on whether EPs of PT-symmetric systems provide an ultimate advantage for sensing applications due to the negative influence of quantum and classical noise, with a fierce debate in the literature. In [230] a lower bound on the sensitivity around an EP operated near the lasing threshold was derived. Despite the large impact of noise on the output, the authors found a large subspace of input states for which the SNR can be enhanced near an EP. They also provided a heterodyne detection scheme based on the obtained bounds, yielding optimal sensitivity in such schemes. In terms of parameter estimation, the sensitivity is usually defined as the minimum parameter change determined above the noise level within a given data acquisition time. In [235], it was shown that the eigenstates' coalescence counteracts the eigenvalue susceptibility divergence and makes the sensitivity a smooth function of the perturbative parameter. Therefore, there is no dramatic enhancement of sensitivity at the EP, but, as in [230], there is a range of parameters and initial conditions for which the sensitivity can be nonetheless enhanced. Sensors operated near EPs are being explored over a broad range of systems, from magnon-cavity systems [236] to optomechanical systems [237], [238]. Although working near an EP can be challenging in experimental realizations, real-time control of the parameters of $H_{\text{eff}}$ for a 3-level superconducting transmon circuit was demonstrated in [239], enabling quantum state tomography of a single dissipative qubit in the vicinity of its EP.

**Conclusions**

In this work, we have presented a tutorial on PT-symmetry and exceptional points in photonics. We have discussed the concepts and ideas that recently emerged in this research area and their potential impact on electromagnetics, antennas and propagation research. First, we have discussed the basic properties of $\hat{P}$, $\hat{T}$ and $\hat{P}\hat{T}$ transformations as applied to quantum mechanics and electromagnetics. Then, we have discussed the main models of PT-symmetric systems and their implications. In the last part of the work, we have also discussed the PT-symmetry and exceptional points in the context of unusual scattering wave phenomena and their applications. We hope that this tutorial may be useful to spur further research activities in this exciting area of research, and enable new application of non-Hermitian electromagnetics.

**Acknowledgment**

Our body of work on these topics has been supported by the Office of Naval Research, the Air Force Office of Scientific Research, the National Science Foundation, the Department of Defense and the Simons Foundation.

**Appendix A**

Let us now define the S-matrix properties in the PT-symmetric scenario in a general two-port system, **Figure 1(a)**. For such a system, we can write the following expressions for the fields on the left (l) and right (r) sides:

$$E_l(x) = i_1 \exp(ik_0 x) + o_1 \exp(-ik_0 x), \text{ for } x < -l, \quad (A1)$$
$$E_r(x) = i_2 \exp(-ik_0 x) + o_2 \exp(ik_0 x), \text{ for } x > l, \quad (A2)$$

where $l$ is the distance between the ports. The action of $\hat{P}$ consists in $x \xrightarrow{\hat{P}} -x$, whereas the action of $\hat{T}$ results in $i \xrightarrow{\hat{T}} -i$, and hence $ik_0 x \xrightarrow{\hat{P}\hat{T}} ik_0 x$, yielding

$$(\hat{P}\hat{T})E_l(x) = (\hat{P}\hat{T})i_1 \exp(ik_0 x) + (\hat{P}\hat{T})o_1 \exp(-ik_0 x), \text{ for } x > l, \quad (A3)$$
$$(\hat{P}\hat{T})E_r(x) = (\hat{P}\hat{T})i_2 \exp(-ik_0 x) + (\hat{P}\hat{T})o_2 \exp(ik_0 x), \text{ for } x < -l. \quad (A4)$$

We see that after the application of the $\hat{P}\hat{T}$ operator, the amplitudes $\hat{P}\hat{T}o_n$ ($\hat{P}\hat{T}i_2$) play the role of incoming (outcoming) fields. These new amplitudes are connected by the same scattering matrix, $\hat{P}\hat{T}i_n = S_{nm}\hat{P}\hat{T}o_m$ and hence $\hat{P}\hat{T}o_m = S_{nm}^{-1}\hat{P}\hat{T}i_n$. On the other hand, acting by $\hat{P}\hat{T}$ on the initial matrix ($o_n = S_{nm}i_m$) and recalling that $(\hat{P}\hat{T})^2 = \hat{1}$, we get $\hat{P}\hat{T}o_n = \hat{P}\hat{T}S_{nm}i_m = (\hat{P}\hat{T}S_{nm}\hat{P}\hat{T})\hat{P}\hat{T}i_m$. Combining the two results, we finally obtain that the S-matrix in PT-symmetric systems is transformed as

$$(\hat{P}\hat{T})S_{nm}(\hat{P}\hat{T}) = S_{nm}^{-1}. \quad (A5)$$